\RequirePackage{fix-cm}
\documentclass[twocolumn]{svjour3}          % twocolumn
\smartqed  % flush right qed marks, e.g. at end of proof
\usepackage{graphicx}
\usepackage{epstopdf}
\usepackage{times}
\usepackage{graphicx}
\usepackage{subfigure}

\usepackage{enumitem}

\usepackage[hyphens]{url}
\usepackage{booktabs} % tables
\usepackage[justification=centering]{caption}

\usepackage{listings} % for source code
\usepackage{xcolor} % for color
\begin{document}

\title{A Parallel Solution to Finding Nodal Neighbors in Generic Meshes%\thanks{Grants or other notes
}


\author{Gang Mei         \and
        Nengxiong Xu     \and
        Hong Tian        \and
        Shengwei Li
        %etc.
}


\institute{G. Mei \and N. Xu \and S. Li\at
              Department of Geological Engineering, Qinghai University, Qinghai, China \\
              School of Engineering and Technolgy, \\China University of Geosciences, Beijing, China \\
              \email{\{gang.mei; xunengxiong; lishengwei\}@cugb.edu.cn}             \\
           \and
           H. Tian \at
           Faculty of Engineering, \\China University of Geosciences, Wuhan, China\\
            \email{htian@cug.edu.cn} \\
}

\date{Received: date / Accepted: date}

\maketitle

\begin{abstract}
In this paper we specifically present a parallel solution to finding the 
one-ring neighboring nodes and elements for each vertex in generic meshes. 
The finding of nodal neighbors is computationally straightforward but 
expensive for large meshes. To improve the efficiency, the parallelism is 
adopted by utilizing the modern Graphics Processing Unit (GPU). The 
presented parallel solution is heavily dependent on the parallel sorting, 
scan, and reduction, and can be applied to determine both the neighboring 
nodes and elements. To evaluate the performance, the parallel solution is 
compared to the corresponding serial solution. Experimental results show 
that: our parallel solution can achieve the speedups of approximately 55 and 
90 over the corresponding serial solution for finding neighboring nodes and 
elements, respectively. Our parallel solution is efficient and easy to 
implement, but requires the allocation of large device memory.
\end{abstract}

\section{Introduction}
\label{sec:introduction}
Mesh generation plays an important role in geometric modeling, computer graphics, and numerical simulations. After generating various types of 
meshes, typically mesh editing is intentionally performed to modify or 
improve the generated meshes to meet desired requirements. In mesh editing 
such as Boolean operations \cite{01_Feito2013705} or mesh optimization \cite{02_DAmato20131127}, the local mesh connectivity especially the adjacent/neighboring nodes and elements for each node or element is frequently needed 
to reduce the computational cost of local search. 

The finding of one-ring nodal neighbors in arbitrary valid mesh is 
computationally straightforward, and can be completely carried out based on 
the connectivity/topology of meshes. The simplest method is to loop over 
all elements in a mesh to identify: (1) which pair of nodes is connected by 
an edge and (2) which nodes are contained in an element \cite{02_DAmato20131127,03_Mei2014,04_Chen2014}. This is 
because that: (1) any pair of nodes connected by an edge is the one-ring 
neighboring node for each other; and (2) any element is directly the 
one-ring neighboring element for those nodes it contains. 

Another simple method for finding the one-ring neighboring nodes for each 
vertex in a polygonal mesh was introduced by Dahal and Newman \cite{05_6950720}. 
They first determined the boundary vertices by finding the opposite edges 
for each vertex and then forming a closed polygon using those opposite 
edges. They adopted the vertices of the closed polygon (i.e., the opposite 
edges) as the one-ring neighbors for each vertex. 

Both of the above neighbors-finding methods are quite easy to implement in 
the serial programming pattern. However, due to the fact that it needs to 
loop over all the elements of a mesh in sequential, the computational cost 
is in general too high for large size of meshes; and this will reduce the 
computational efficiency of the entire mesh editing procedure. 

An effective strategy to improve the efficiency of the neighbors-finding 
procedure is to parallelize it on various parallel computing architectures 
such as multi-core CPUs or many-core GPUs, i.e., to design parallel solution 
to finding neighbors. 

However, when finding nodal neighbors in parallel, there exists the race 
condition. The race condition issue appears when two different parallel 
threads may need to be written in the same memory position \cite{06}. When looping over all the elements in a mesh to find 
the nodal neighbors, two neighboring nodes of the same vertex may be found 
concurrently within two parallel threads; and the indices of the two 
neighboring nodes may need to be written in the same memory position for 
storing. In this case, race condition arises. 

Currently, the most commonly used method to address the above problem is to 
color the mesh first and then looping over those elements with the same 
color simultaneously to find neighbors \cite{04_Chen2014,06,07,08_Benitez2014,09_Cheng2015}. This coloring-based 
method is very effective and efficient for large-scale meshes, and is quite 
suitable to be applied in parallel pattern. The only minor shortcoming is 
that: it is needed to color the mesh into several groups of elements and 
thus needs additional computational cost. 

In this paper, without the use of mesh-coloring, we specifically design and 
develop a parallel solution to finding nodal neighbors by utilizing the 
power of modern GPUs. Our solution is efficient, simple and easy to 
implement, which heavily depended on the use of parallel primitives such as 
sorting, scan, and reduction. In addition, in our solution there is no need 
to adopt complex mesh data structures; and only arrays of integers are 
required. To evaluate the performance of our parallel solution, we compare 
it to the corresponding serial solution in several experiments. 

\section{Methods}

\subsection{Basic Ideas behind Our Solution}

\subsubsection{The Idea for Finding Neighboring Nodes for Each Vertex}

\begin{figure*}[htbp]
	\centering
	\subfigure[]{
		\label{fig:1a}       % Give a unique label
		\includegraphics[scale = 0.65]{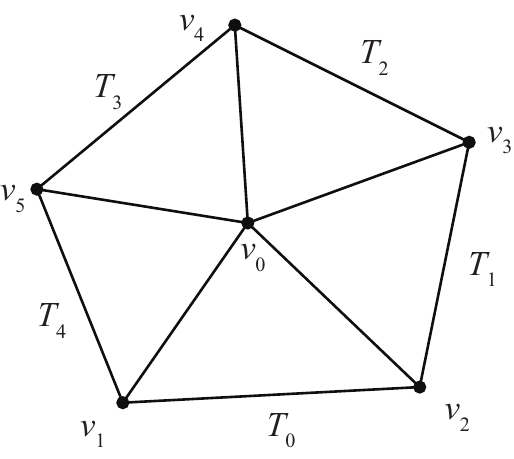}
	}
	\hspace{1em}
	\subfigure[]{
		\label{fig:1b}       % Give a unique label
		\includegraphics[scale = 0.65]{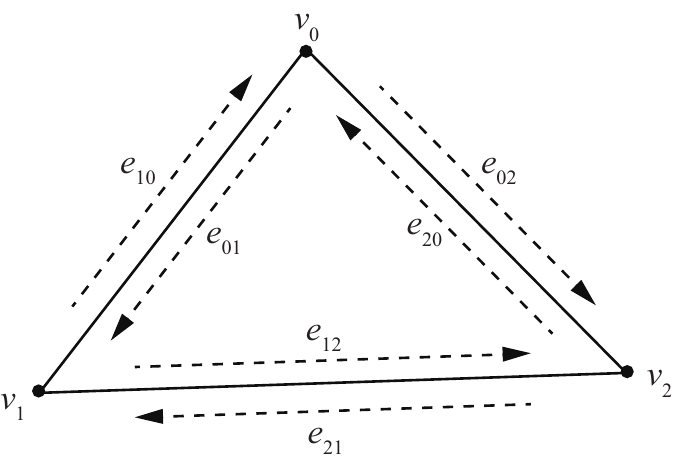}
	}
	\hspace{1em}
	\subfigure[]{
		\label{fig:1c}       % Give a unique label
		\includegraphics[scale = 0.65]{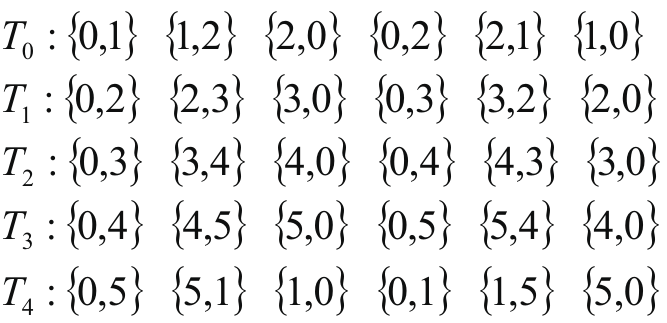}
	}
	\hspace{1em}
	\subfigure[]{
		\label{fig:1d}       % Give a unique label
		\includegraphics[scale = 0.65]{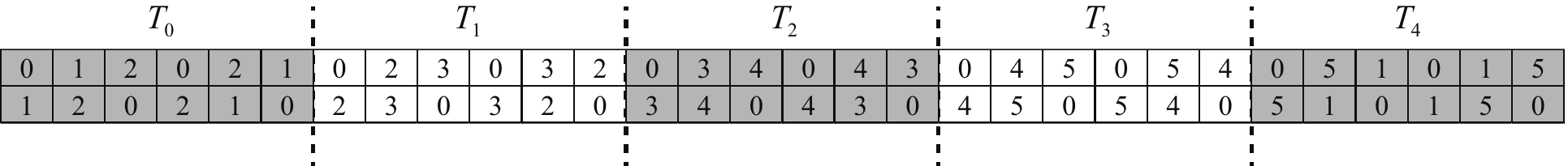}
	}
	\hspace{1em}
	\subfigure[]{
		\label{fig:1e}       % Give a unique label
		\includegraphics[scale = 0.65]{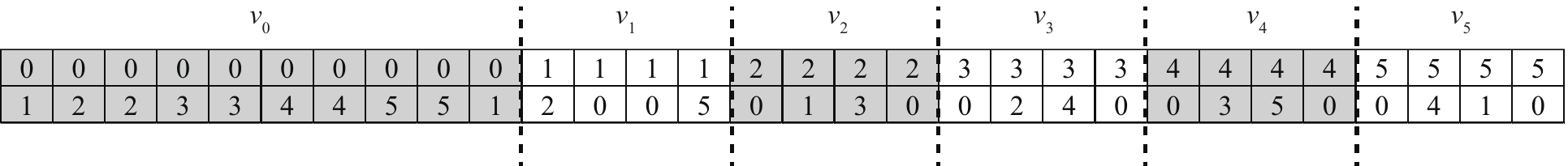}
	}
	\caption{A simple illustration of finding neighboring nodes for each vertex in a mesh}
	\label{fig:1}       % Give a unique label
\end{figure*}

In any valid meshes, any pair of neighboring nodes is connected using an 
edge. An edge can be represented with two nodes (and further, the indices of 
two nodes). A mesh typically has plenty of edges (i.e., pairs of nodes). 
When gathering all edges of a mesh (see Figure \ref{fig:1a} and \ref{fig:1b}), the edges can be stored in an array 
consisting of $n$ pairs of integer values; see Figure \ref{fig:1c}. The array of edges 
can be also considered as two arrays of integers; see Figure \ref{fig:1d}. If 
adopting the first array of integers as the keys for sorting and the second 
array and the correspondingly attached values, the indices of all the 
neighboring nodes for the same vertex can be easily found by performing a 
parallel sorting; see Figure \ref{fig:1e}.

\subsubsection{The Idea for Finding Neighboring Elements for Each Vertex}

\begin{figure*}[htbp]
	\centering
	\subfigure[]{
		\label{fig:2a}       % Give a unique label
\includegraphics[scale = 0.65]{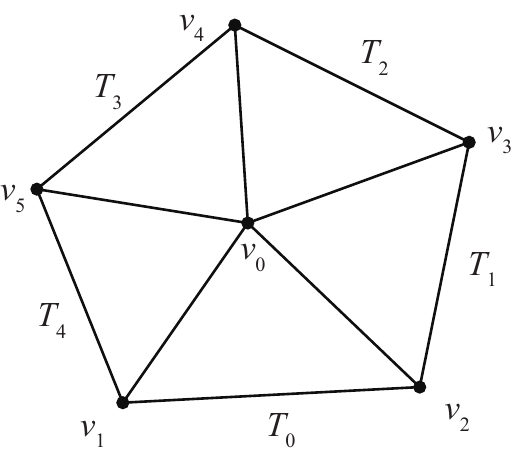}
	}
	\hspace{1em}
	\subfigure[]{
		\label{fig:2b}       % Give a unique label
\includegraphics[scale = 0.65]{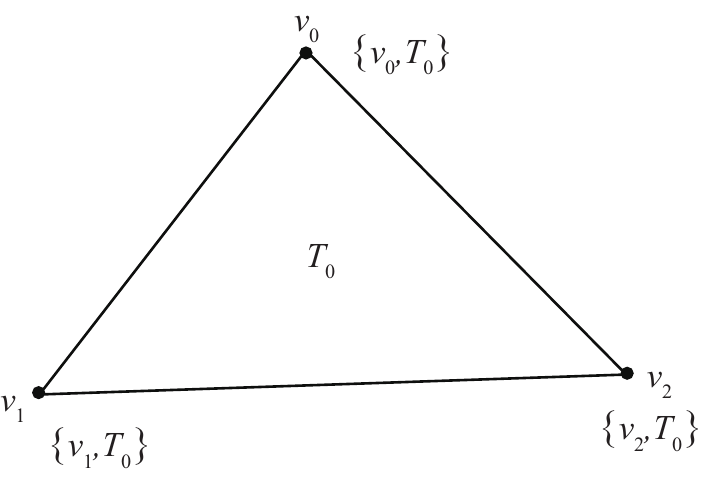}
	}
	\hspace{1em}
	\subfigure[]{
		\label{fig:2c}       % Give a unique label
\includegraphics[scale = 0.65]{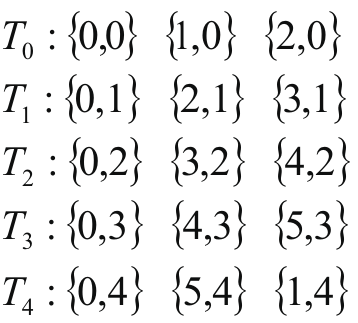}
	}
	\hspace{1em}
	\subfigure[]{
		\label{fig:2d}       % Give a unique label
\includegraphics[scale = 0.65]{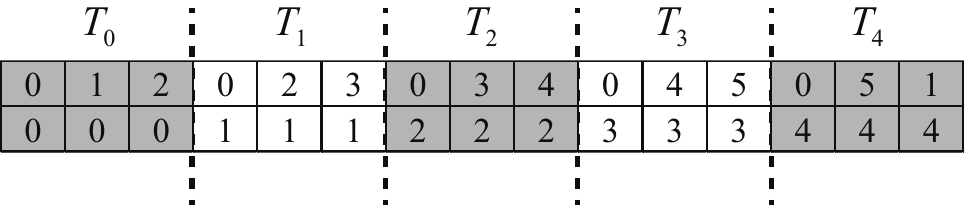}
	}
	\hspace{1em}
	\subfigure[]{
		\label{fig:2e}       % Give a unique label
\includegraphics[scale = 0.65]{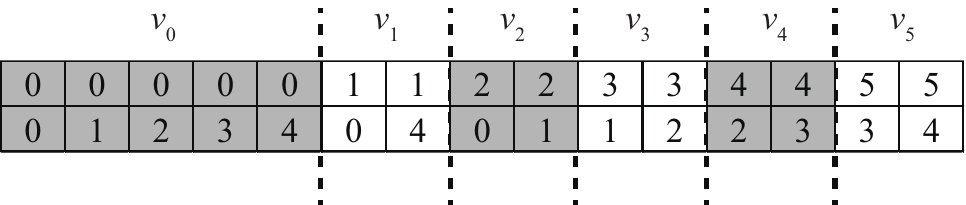}
	}
	\caption{A simple illustration of finding neighboring elements for each vertex in a mesh}
	\label{fig:2}       % Give a unique label
\end{figure*}

An element in a mesh is composed of several nodes (Figure \ref{fig:2a}). Each element is by nature 
the one-ring neighboring element of those nodes it contains. A pair of 
integer values can be used to simply demonstrate this relationship: the 
first integer is the index of one of the nodes in an element; and the second 
is the index of the element itself; see Figure \ref{fig:2b}. 

For an element, several such pairs of integers can be formed. And for an 
arbitrary mesh, a group of such pairs of integers can be obtained, and 
stored in two arrays of integers. Similar to the finding of neighboring 
nodes for each vertex, if adopting the first array of integers as the keys 
for sorting and the second array of integers as the correspondingly attached 
values, the indices of all the neighboring elements for the same vertex can 
be easily found by performing a parallel sorting; see Figure \ref{fig:2}.

\subsection{Implementation Details}

Our solution is applicable to arbitrary meshes. However, to demonstrate our 
solution, here we specifically present the implementation details of our 
solution when applied to the triangular surface mesh. 

Our solution is implemented by strongly utilizing the parallel primitives 
provide by the library \textit{Thrust} \cite{10_Bell2012359} such as the parallel sorting and 
reduction. Thrust is a parallel algorithms library containing the efficient 
parallel primitives such parallel scan, reduction, and sorting. In our 
solution, the quite efficient primitive, parallel sorting is adopted to 
extremely fast sort the arrays of integers to find the neighboring nodes and 
elements. More implementation details are listed as follows.

\subsubsection{The Finding of Neighboring Nodes}

The process of finding the neighboring nodes for each vertex is as follows.

\begin{figure*}[htbp]
	\centering
    \includegraphics[scale = 0.75]{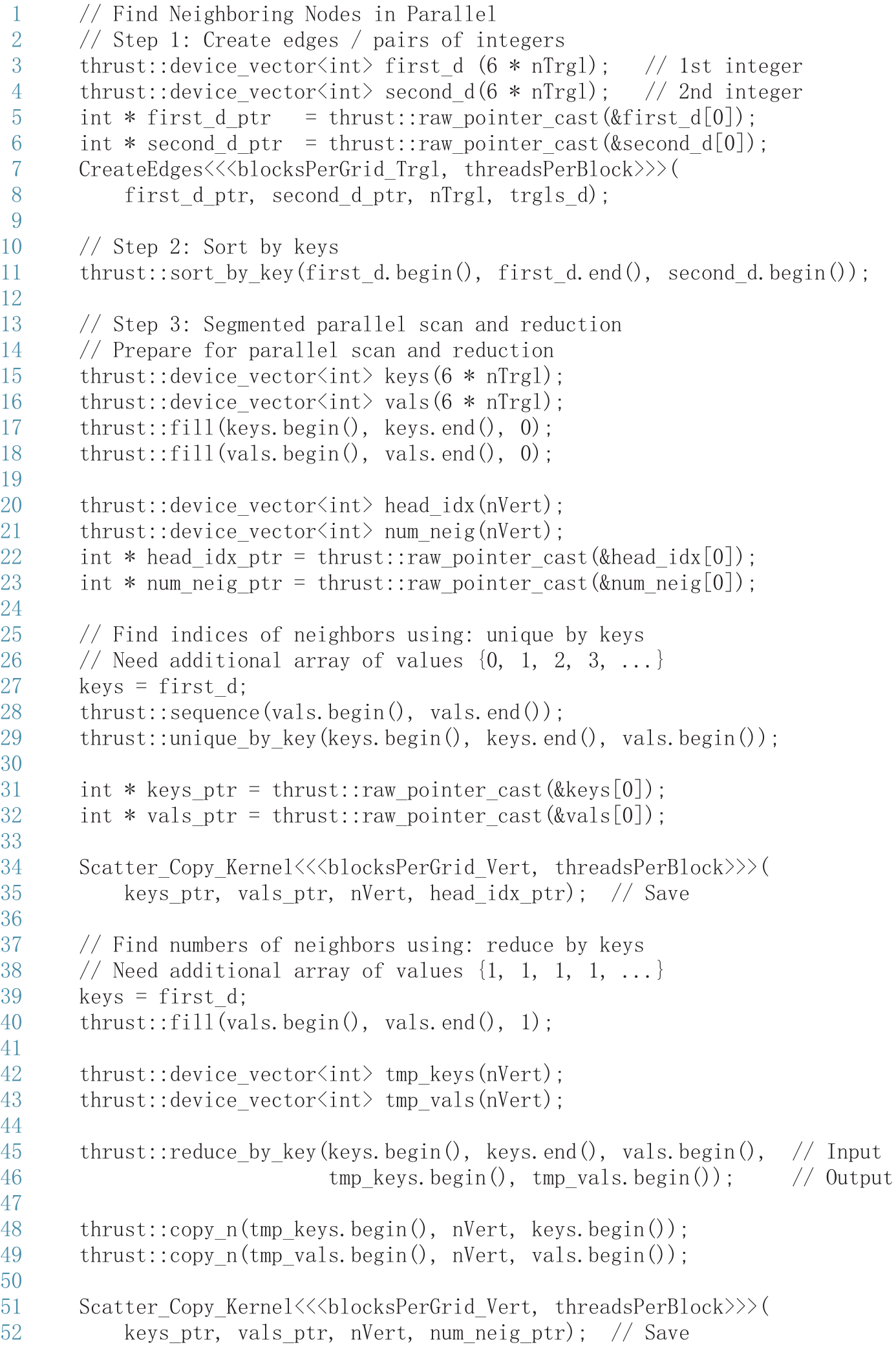}
	\caption{Partial GPU implementation for finding neighboring nodes}
	\label{fig:code}       % Give a unique label
\end{figure*}

First, we create pairs of integers according to the edges in each triangle. 
Noticeably, we can form three edges (pairs of integers) for a triangle when 
the three nodes in the triangle are organized in count-clockwise (CCW) order 
and another three pairs when nodes are organized in clockwise (CW); see 
Figure \ref{fig:1b}. That is, a triangle can produce six pairs of integers. These 
edges/pairs can be obviously created in parallel. We specifically design a 
CUDA kernel to realize this; see the lines 7 $\sim $ 8 in Figure \ref{fig:code}.

After creating the pairs of integers that are stored in two arrays of 
integers, the second step is to sort those pairs according to the first 
array of integers (see line 11 in Figure \ref{fig:code}). This procedure can be 
extremely fast performed by using the specific function \texttt{thrust:: 
	sort{\_}by{\_}keys()}. 

The third step is to determine (1) the total number and (2) the 
detailed indices of the neighboring nodes for each vertex, which can be 
realized by using segmented scan and reduction. The ideas behind performing 
the segmented scan and reduction have been presented in our previous work \cite{11_Mei2016}.

To determine the number of neighbors, we first create a helper array 
containing the same value 1 (i.e., line 40 in Figure \ref{fig:code}), and then perform a 
parallel segmented reduction by using the function 
\texttt{thrust::reduce{\_}by{\_}keys()}; see lines 42 $\sim $ 46 in Figure \ref{fig:code}. To obtain the indices of the neighboring nodes, we also first create a 
helper array of sequenced integers (i.e., line 28 in Figure \ref{fig:code}), and then 
perform a parallel segmented scan by using  
\texttt{thrust::unique{\_}by{\_}keys()}; see lines 27 $\sim $ 29 in Figure \ref{fig:code}. After performing the segmented reduction and scan, both the number and 
indices of neighbors can be found and then transferred into other target 
arrays for further mesh editing such as Boolean operations or mesh 
optimization. 

\subsubsection{The Finding of Neighboring Elements}

The process of finding the neighboring elements is quite similar to that of 
finding the neighboring nodes. The first step is also to form the pairs of 
integers (i.e., two arrays of integers), then to sort according to the first 
array of integers, and third use the parallel segmented reduction and scan 
to further determine both the total number and the detailed indices of the 
neighboring elements. 

However, there is a remarkable difference between the process of finding 
neighboring nodes and elements. When finding the neighboring elements, the 
first integer value of any pair is the index of a node in an element; and 
the second value of the pair of integers is the index of the element itself. 
In contrast, in the finding of neighboring nodes, both the first and the 
second integer value of any pair is the index of node. 

\section{Results and Discussion}

In this section, we will evaluate the performance of our parallel solution 
by comparing to the corresponding serial solution when applied to the 
triangular surface meshes. Moreover, we will analyze the advantages and 
shortcomings of our parallel solution based on the experimental results. 

\subsection{Results}

Six groups of experimental tests are carried out to evaluate the performance 
of our parallel solution. These experimental tests of the parallel solution 
are performed on the desktop computer which features with the NVIDIA GeForce 
GT640 (GDDR5) graphics card with 1GB memory and the GPU programming model 
CUDA v7.0. The experiments of the corresponding serial solution are 
performed on Windows 7 SP1 with an Intel i5-3470 CPU (3.2 GHz and 4 Cores) 
and 8GB of RAM memory. 

\begin{figure*}[htbp]
	\centering
    \includegraphics[scale = 0.55]{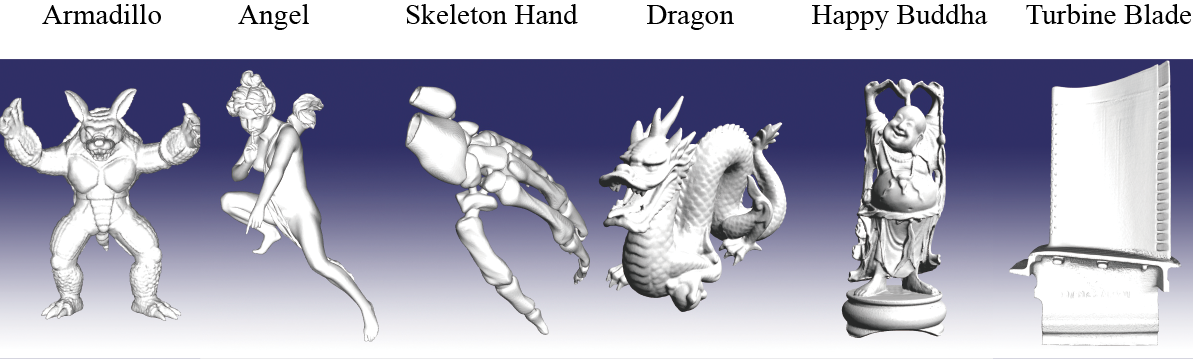}
	\caption{Six mesh models that are employed for experimental tests}
	\label{fig:mesh:model}       % Give a unique label
\end{figure*}

These six triangular surface mesh models employed for testing are directly 
obtained from the Stanford 3D Scanning Repository (\url{http://www.graphics.stanford.edu/data/3Dscanrep/}) 
and the GIT Large Geometry Models Archive (\url{http://www.cc.gatech.edu/projects/large_models/}); 
see Figure \ref{fig:mesh:model} and Table 
\ref{tab1}. The execution time of both our parallel solution and the serial solution 
for finding the neighboring nodes and elements are listed in Table \ref{tab1}.

The experimental results listed in Table \ref{tab1} indicate that: (1) our parallel 
solution is approximately 55 and 90 times faster than the  
serial solution when finding the neighboring nodes and elements, 
respectively; and (2) for the entire neighbors-finding procedure, our 
parallel solution achieves the speedup of approximately 60 over the serial 
solution.

\begin{table*}[htbp]
	\caption{Comparison of efficiency of our parallel solution and the serial solution (1k = 1000)}
	\centering
	\begin{tabular}
		{|c|p{31pt}|p{32pt}|p{21pt}|p{21pt}|p{21pt}|p{21pt}|p{21pt}|p{21pt}|p{21pt}|}
		%{|p{30pt}|p{31pt}|p{36pt}|p{21pt}|p{21pt}|p{21pt}|p{21pt}|p{21pt}|p{21pt}|p{21pt}|}
		\hline
		{Mesh Model}& 
		{Num. of Vertex}& 
		{Num. of Triangle}& 
		\multicolumn{2}{|c|}{Serial Time (/ms)} & 
		\multicolumn{2}{|c|}{Parallel Time (/ms)} & 
		\multicolumn{3}{|c|}{Speedup}  \\
		\cline{4-10} 
		& 
		& 
		& 
		Finding Node& 
		Finding Elem.& 
		Finding Node& 
		Finding Elem.& 
		Finding Node& 
		Finding Elem.& 
		Overall \\
		\hline
		Armadillo& 
		172k& 
		346k& 
		2527& 
		1732& 
		50.0& 
		22.1& 
		50.5& 
		78.4& 
		59.1 \\
		\hline
		Angel& 
		237k& 
		474k& 
		3635& 
		2580& 
		66.7& 
		29.4& 
		54.5& 
		87.8& 
		64.7 \\
		\hline
		Skeleton Hand& 
		327k& 
		655k& 
		4806& 
		3305& 
		89.1& 
		38.3& 
		53.9& 
		86.3& 
		63.7 \\
		\hline
		Dragon& 
		437k& 
		871k& 
		6543& 
		4488& 
		112.7& 
		48.5& 
		58.1& 
		92.5& 
		68.4 \\
		\hline
		Happy Buddha& 
		543k& 
		1088k& 
		8577& 
		6072& 
		138.5& 
		60.6& 
		61.9& 
		100.2& 
		73.6 \\
		\hline
		Turbine Blade& 
		882k& 
		1765k& 
		13931& 
		10000& 
		221.3& 
		93.3& 
		63.0& 
		107.2& 
		76.1 \\
		\hline
	\end{tabular}
	\label{tab1}
\end{table*}

\subsection{Discussion}

The finding of neighboring nodes and elements for each vertex in arbitrary 
meshes is computationally straightforward, but expensive for large meshes. 
In this paper, we specifically design and develop a parallel solution to 
improve the computational efficiency. In this section, we will analyze the 
advantages and shortcomings of our parallel solution.

\subsubsection{The Advantages of Our Parallel Solution}

The first advantage of our parallel solution is that: it is easy to 
implement due to the reason it does not need to perform the mesh-coloring 
before finding neighbors. The mesh-coloring technique is frequently used to 
deal with the race condition issue \cite{06}. 

When finding nodal neighbors in parallel, there exists the race condition. 
The race condition issue appears when two different parallel threads may 
need to be written in the same memory position \cite{06}. When 
looping over all the elements in a mesh to find the nodal neighbors, two 
neighboring nodes of the same vertex may be found concurrently within two 
parallel threads; and the indices of the two neighboring nodes may need to 
be written in the same memory position for storing. In this case, race 
condition arises. 

Currently, the most commonly used method to address the above problem is to 
color the mesh first and then looping over those elements with the same 
color simultaneously to find neighbors \cite{04_Chen2014,06,07,08_Benitez2014,09_Cheng2015}. This coloring-based 
method is very effective and efficient for large-scale of meshes, and is 
quite suitable to be applied in parallel pattern. The only minor shortcoming 
is that: it is needed to color the mesh into several groups of elements and 
thus needs additional computational cost. 

In our parallel solution, we have redesigned the process of finding 
neighbors to avoid the use of mesh-coloring by strongly exploiting those 
efficient parallel primitives such as parallel sorting and scan. In 
addition, there are no complex data structures; and only arrays of integers 
are needed. Thus, our parallel solution is easy to implement in practice. 

The second advantage of our parallel solution is the acceptable efficiency. 
The experimental results listed in Table \ref{tab1} indicate that: our parallel 
solution can generate the speedups of approximately 55 and 90 over the 
serial solution when finding the neighboring nodes and elements, 
respectively. 

This performance gains benefit from the parallelization performed on the 
GPU. By analyzing and reorganizing the process of finding nodal neighbors, 
we have transferred the entire process into several sub-procedures of 
parallel sorting, scan, and reduction, while these parallel primitives are 
extremely fast for the large size of data.

Another reason why our parallel solution is quite efficient is that: there 
are no complex data structures; and only arrays of integers are needed. 
Inherently, operations and computations for discrete arrays of integer 
values rather than arrays of structures are quite fast on the GPU. We 
specifically avoid using arrays of structures such as pairs, but chose to 
directly use the arrays of values. This leads additional performance gains.

\subsubsection{The Shortcomings of Our Parallel Solution}

Although our parallel solution is efficient and easy to implement, it has an 
obvious shortcoming, i.e., it requires more device memory than that 
of the serial solution. This additionally required device memory is 
allocated for performing the sorting, scan, and reduction. 

In the serial solution, a STL (C++ Standard Template Library) container, 
\texttt{vector<int>}, is adopted to allocate an array to dynamically store 
the indices of neighboring nodes for each vertex. The size of the array can 
be dynamically determined without redundant space. Similar, another array of 
integers is needed to hold the indices of neighboring elements. Moreover, 
the number of neighboring nodes and elements are directly the size of the 
dynamic arrays, which can be easily and automatically determined. Thus, 
there is no need to allocate other additional arrays. 

In contrast, in our parallel solution, six additional arrays of integers are 
required. First, two arrays of integers need to be allocated to store the 
edges (i.e., pairs of integers). Second, another two arrays of integers are 
needed to hold the first indices and numbers of neighboring nodes. And 
third, to perform the segmented parallel reduction and scan, another two 
temporary arrays are also required. 

Due to the required additional arrays of integers, our parallel solution 
cannot be applied to quite large size of meshes since the device memory (the 
global memory) of most current GPUs is quite limited. Thus, future work may 
focus on redesigning the parallel process of finding nodal neighbors to 
reduce the device memory. 

\section{Conclusions and Outlook}

In this paper, we have designed and developed a parallel solution to finding 
the neighboring nodes and elements for each vertex in an arbitrary mesh by 
exploiting the GPU. Our solution is a topology-based method, and is heavily 
dependent on the use of parallel sorting, scan, and reduction. We have 
compared our parallel solution to the corresponding serial solution to 
evaluate its performance. We have found that: our parallel solution is 
approximately 55 and 90 times faster than the corresponding serial solution 
when finding the neighboring nodes and elements, respectively. Our solution 
is efficient, simple and easy to implement, and can be applied for arbitrary 
meshes. However, our parallel solution requires the allocation of large 
device memory; and thus future work is planned to be carried out to address 
this problem. To benefit the community, the complete source code of our 
solution is publicly available for any potential usages. 

%% For one-column wide figures use
%\begin{figure}
%% Use the relevant command to insert your figure file.
%% For example, with the graphicx package use
%  \includegraphics{example.eps}
%\end{figure}
%%
%% For two-column wide figures use
%\begin{figure*}
%% Use the relevant command to insert your figure file.
%% For example, with the graphicx package use
%  \includegraphics[width=0.75\textwidth]{example.eps}
%\end{figure*}
%\begin{table}
%\begin{tabular}{lll}
%\hline\noalign{\smallskip}
%\end{tabular}
%\end{table}

\begin{acknowledgements}
This research was supported by the Natural Science Foundation of China 
(Grant Numbers: 40602037, 40872183, and 51541405), China Postdoctoral Science Foundation 
(2015M571081), and the Fundamental Research Funds for China Central 
Universities (2652015065). The authors would like to thank the editor and 
the reviewers for their contributions on the paper. 
\end{acknowledgements}

\bibliographystyle{spphys}       % APS-like style for physics
\bibliography{GM}   % name your BibTeX data base

\end{document}